# LYMAN ALPHA FOREST CLUSTERING USING CELL COUNTS STATISTICS

François R. Bouchet[1], and Avery Meiksin[2]
[1] *Institut d'Astrophysique de Paris, CNRS, Paris, France.*
[2] *University of Chicago, Department of Astronomy & Astrophysics, Chicago, USA.*

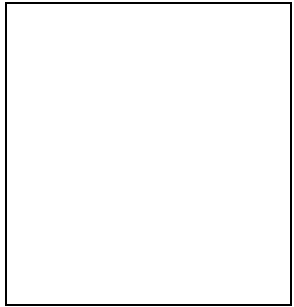

**Abstract**

We present a novel technique for calculating Ly$\alpha$ forest correlations based on cell counts. It is applied to the line lists from 7 QSOs observed at high resolution ($\Delta v < 25\,\mathrm{km\,s^{-1}}$). Two spectra (Q0055−259 and Q0014+813) appear to be sufficiently free of biases to obtain meaningful estimates. We find for both positive correlations, with a maximum of $0.5-1$, on the scale of $0.5-3 h^{-1}$ Mpc (comoving), or $100-600\,\mathrm{km\,s^{-1}}$, which is consistent with the primordial power spectrum inferred from optically-selected galaxy surveys. Strong evidence for anti-correlation on the scale of $3-6 h^{-1}$ Mpc is also found. If this first detection of anti-correlations is physical, additional factors to gravity may play a role in determining the clustering of the Ly$\alpha$ forest on comoving scales $> 3h^{-1}$ Mpc.

Attempts to search for correlations in the Ly$\alpha$ forest have generally led to inconclusive results (e.g., Webb 1986; Rauch et al. 1992). Recently, the results of a high spectral resolution observation of Q0055−269 by Cristiani et al. (1995) has produced the strongest evidence yet for clustering in the Ly$\alpha$ forest. They report an amplitude of $\sim 0.5$ on the scale of a few hundred kilometers per second (see also Cristiani, this volume). Here we describe a novel method whose results confirm earlier findings, and provide the first evidence for anti-correlations at larger scales.

A homogeneous and isotropic statistical point process may be completely described by the correlation functions of the system. The distribution functions of the neighbor counts (NPDF) is an equivalent description, related to the former through a set of infinite sums (White 1979). In particular, the two-point correlation function $\xi(s)$, where $s$ is a measure of the separation of the objects, may be expressed as the sum over the NPDF, $P_N$, the distribution probabilities of finding the $N+1$ nearest neighbors a distance $s$ from a given point: $n\left[1 + \xi(s)\right]\,ds = n\sum_{N=0}^{\infty} P_N(s) ds$ where $n$ is the number density of objects per unit separation $s$. For a Poisson process, the NPDF is $\hat{P}_N(s) = \mu^N \exp(-\mu)/N!$, where $\mu = ns$. It is convenient to multiply the previous equation by the density of objects, integrate over all space, and subtract off the Poisson contribution, to relate the integral of the two-point function to the cumulative distributions $C_N$ of neighbor counts. Denoting by $dN/dz$ the density of absorbers per unit redshift, and by $Z$ the redshift integration range, the result is

$$\int_Z dz \left(\frac{dN}{dz}\right)^2 \frac{dz}{ds} \int_{s_{\min}}^{s} ds' \xi(s') = \sum_{N=0}^{\infty} \left[C_N(s) - \hat{C}_N(s)\right], \qquad (1)$$

where $\hat{C}_N(s)$ is the cumulative distribution of neighbor counts expected for a Poisson process, $\hat{C}_N(s) = \int_Z dz \left(\frac{dN}{dz}\right)^2 \int_{s_{\min}}^{s} ds' \frac{dz}{ds'} \hat{P}_N(s')$. This form has the useful feature that the maximum difference between $C_N(s)$ and $\hat{C}_N(s)$ (normalized to unity for $s \to \infty$), is the Kolmogorov-Smirnov measure of distance between the two distributions. It provides a direct estimate for the probability that the statistics of the distribution are non-Poissonian. The minimum separation $s_{\min}$ in eqs.(1) accounts for the finite resolution of the QSO spectra, set according to the minimum resolvable absorption line separation as defined by the observers. This procedure indeed results in a matching between the observed and predicted cell size distributions for small separations. To extract the correlation function from its integral, we neglect any dependence on redshift, since the Ly$\alpha$ forest in a given QSO is measured only over a small redshift interval. The integrals may then be reversed in eq.(1), yielding an integral of $\xi(s)$ only over $s$. The correlation function is then obtained by computing a smoothed derivative of the integral, using the kernel $M_4(x) = 2/3 - x^2 + |x|^3/2$ for $0 \leq x \leq 1$, $(2-|x|)^3/6$ for $1 \leq x \leq 2$, and 0 for $x \geq 2$, where $x = s/h$, and $h$ is the smoothing length (e.g., Monaghan 1985). We checked on realistic test problems that this approach converges (from above) to the correct correlation function as the density of lines is increased.

We estimated the correlation functions for the Ly$\alpha$ forests measured in 7 QSOs observed at high resolution ($< 25\,\mathrm{km\,s^{-1}}$), over the redshift range $2 < z < 4$. The results for two of these, Q0055–269 at $z = 3.66$ (Cristiani et al. 1995), and Q0014+813 at $z = 3.38$ (Rauch et al. 1992), are shown in Figure 1a. The correlation functions are computed as a function of comoving separation $r_c$, for $H_0 = 100h\,\mathrm{km\,s^{-1}\,Mpc^{-1}}$. The conversion to velocity separation is given by $\Delta v \simeq 200\Delta r_c$. Only systems with reported column densities exceeding $10^{13.75}\,\mathrm{cm^{-2}}$ are included. Because metal systems show positive correlations (e.g., Young et al. 1982), we exclude from consideration the lines within $\pm 500\,\mathrm{km\,s^{-1}}$ (in the local rest frame) of the metal systems quoted by the observers[1]. To avoid the proximity effect (Murdoch et al. 1986), we also exclude the region within $7000\,\mathrm{km\,s^{-1}}$ of the QSO redshift. The smoothing length used to extract $\xi(s)$ from its integral is $50\,\mathrm{km\,s^{-1}}$. The results quoted below for the integral of $\xi(s)$ are independent of this choice.

In agreement with Cristiani et al. , we find evidence for positive correlations on scales up to $3h^{-1}\,\mathrm{Mpc}$, or $\sim 600\,\mathrm{km\,s^{-1}}$, with $\xi \simeq 0.5 - 1$ on the scale of $0.5h^{-1}\,\mathrm{Mpc}$. The integrated correlation function between 0.5 and 1.5 $h^{-1}$ Mpc is 1.02, and between 0.5 and 3 $h^{-1}$ Mpc is 1.46. The number density of lines found is consistent with an average density of $dN/dz = 4.96(1+z)^{2.46}$, where the exponent is adopted from the analysis of Press et al. (1993). Generating 1000 Monte Carlo simulations with this line density, we find that the probabilities of obtaining values as large as those measured in these two intervals are 0.3% and 5.7%, respectively. We also find evidence for anticorrelation on the scale of $3 - 6h^{-1}$ Mpc. The integrated correlation function over this range is $-0.83$, with a Poisson probability for a value this low of $0.6\%$.

---
[1] We also check for possible contamination by C IV doublets misidentified as Ly$\alpha$ lines in the spectra of Q0055–269 and Q0014+813, for systems with an inferred H I column density exceeding $10^{13.3}\,\mathrm{cm^{-2}}$. We find only one candidate C IV system, at $z = 2.5642$ in the spectrum of Q0055–269, but with quoted H I column densities smaller than $10^{13.75}\,\mathrm{cm^{-2}}$.

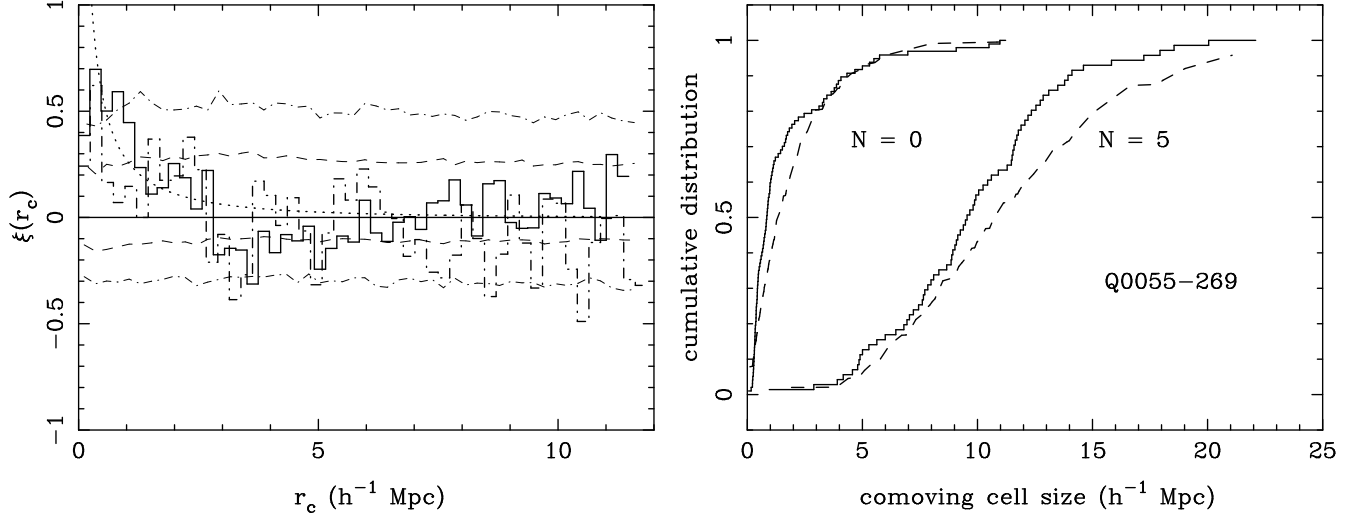

Figure 1: (a) Cell count estimate of $\xi(r_c)$, for Q0055−269 (*solid*), and Q0014+813 (*dot-dashed*). The error ranges shown are the $1\sigma$ and $2\sigma$ variances for the corresponding Poisson distribution. The correlation function for unbiased $\Omega = 1$ CDM with $h = 0.5$ and $\sigma_8 = 1$ (assuming linear growth, $\xi \propto (1+z)^{-2}$, *dotted line*) is consistent with the clustering of galaxies on scales smaller than $20h^{-1}$ Mpc. (b) Distribution of cell sizes for $N = 0$ (gaps) and $N = 5$ neighbors, compared with the expectations for a random distribution of lines, for Q0055−269.

An even stronger measure of the deviation from a Poisson distribution for the absorber positions is provided by the distribution of cell sizes for different cell counts. We show the measured and predicted cell size distributions for cell counts $N = 0$ (gaps) and $N = 5$ in Figure 1b. The estimated distribution is based on assuming a minimum separation of $14 \, \text{km s}^{-1}$, the reported resolution of the spectrum. The measured distribution matches the predicted for small and large separations. The maximum difference between the measured and predicted gap distribution is $0.222$. The KS probability of obtaining so large a difference is $1.4 \times 10^{-4}$. The void distribution yields weaker evidence for clustering in the range $13.3 < \log N_{\text{HI}} < 13.8$, with a KS probability of $7.0\%$ for the distribution to be Poisson. Marginal evidence is found for a non-Poisson distribution over the range $13.3 < \log N_{\text{HI}} < 13.6$, with a KS probability that the void distribution is Poisson of $11\%$. The pair count estimate of Cristiani et al. yielded no evidence for clustering in this range.

Positive, somewhat weaker, correlations over the same scale are found for the forest measured in Q0014+813 as well, although the level of significance is much smaller. The integrated correlation function over the range $0.5 - 3h^{-1}$ Mpc is $0.97$, the probability for which we find to be $17\%$, adopting a line density of $5.05(1+z)^{2.46}$ to match the observed number of lines in this QSO. Intriguingly, negative correlations are again found over the range $3 - 6h^{-1}$ Mpc. The integrated correlation function over this range is $-0.41$, with a probability of $13\%$ for obtaining so low a value. The KS test applied to the void distribution provides stronger evidence for clustering, rejecting a Poisson distribution at the $97.6\%$ confidence level.

We have measured the correlation functions of the Ly$\alpha$ forest in 5 additional QSOs with spectra observed at better than $25 \, \text{km s}^{-1}$ resolution, Q2126−158 (Giallongo et al. 1993), Q1946+7658 (Fan & Tytler 1994), Q2206−199N (Pettini et al. 1990; Rauch et al. 1993), Q1100−264 (Carswell et al. 1991), and Q1331+170 (Kulkarni et al. 1995). We do not claim a detection of physical correlations in these spectra, however. While strong, highly significant correlations are found, with a typical amplitude of $2-3$ at a comoving separation of $0.5-1 h^{-1}$ Mpc, we find large positive and negative excursions at larger separations which greatly exceed the random expectations. While these excursions may in principle be revealing large-scale inhomogeneities in the Ly$\alpha$ forest, we suspect they represent instead the interplay between the varying level of noise across the spectra and the algorithms employed to detect and measure the individual absorption features. Reobserving these spectra at higher signal-to-noise, with comparable or improved resolution, is required to decide the reality of the correlations. In one additional case examined, Q1442+101 at $z = 3.54$ (Frye et al. 1994), the correlations found may directly be attributed to the line-finding algorithm.

In summary, for comoving separations less than $3 h^{-1}$ Mpc, remarkably good agreement is found between the measurements and the expected correlations deduced using the measured galaxy power spectrum, suggesting that the Ly$\alpha$ forest originates from the same primordial density fluctuations as do the galaxies. If the anticorrelations on larger scales are physical, they suggest non-gravitational effects are at work on scales exceeding $3 h^{-1}$ Mpc. A plausible candidate for modifying the correlation function is the UV photoionizing background.